\documentclass[twocolumn,english,reprint]{revtex4}
\usepackage[T1]{fontenc}
\usepackage[latin9]{inputenc}
\usepackage{amsthm}
\usepackage{amsmath}
\usepackage{graphicx}

\makeatletter

\newcommand{\lyxmathsym}[1]{\ifmmode\begingroup\def\b@ld{bold}
  \text{\ifx\math@version\b@ld\bfseries\fi#1}\endgroup\else#1\fi}

\@ifundefined{textcolor}{}
{%
 \definecolor{BLACK}{gray}{0}
 \definecolor{WHITE}{gray}{1}
 \definecolor{RED}{rgb}{1,0,0}
 \definecolor{GREEN}{rgb}{0,1,0}
 \definecolor{BLUE}{rgb}{0,0,1}
 \definecolor{CYAN}{cmyk}{1,0,0,0}
 \definecolor{MAGENTA}{cmyk}{0,1,0,0}
 \definecolor{YELLOW}{cmyk}{0,0,1,0}
 }

\makeatother

\usepackage{babel}
\begin{document}

\title{High temperature structural and magnetic properties of cobalt nanorods.}

\author{Kahina Ait Atmane$^{a}$, Fatih Zighem$^{b}$, Yaghoub Soumare$^{a}$, Mona Ibrahim$^{c}$,
Rym Boubekri$^{c}$, Thomas Maurer$^{b}$, Jérémie Margueritat$^{b}$,
Jean-Yves Piquemal$^{a}$, Frédéric Ott$^{b}$, Grégory
Chaboussant$^{b}$, Frédéric Schoenstein$^{d}$ Noureddine
Jouini$^{d}$ and Guillaume Viau$^{c}$}

\affiliation{$^{a}$ITODYS, CNRS/Université Paris Diderot, Sorbonne Paris Cité,
75205 Paris, France}

\affiliation{$^{b}$LLB, CEA/CNRS, IRAMIS, 91191 Gif sur Yvette, France}

\affiliation{$^{c}$LPCNO, CNRS/INSA-Université de Toulouse, 31077 Toulouse, France}

\affiliation{$^{d}$ LSPM, CNRS/Université Paris XIII, 93430 Villetaneuse, France}

\email{jean-yves.piquemal@univ-paris-diderot.fr (J.-Y.P.); gviau@insa-toulouse.fr (G.V.)}

\selectlanguage{english}%
\begin{abstract}

\end{abstract}
\maketitle

\section{Introduction}

The last decade has seen numerous investigations on anisotropic inorganic
nanoparticles such as nanowires (NWs) and nanorods (NRs) for their
peculiar physical properties and new applications in optics, magnetism
or electronics \cite{1_lieber_2007,2_xia_2003}. Several physical
and chemical methods have been developed to grow ferromagnetic nanowires.
The most used consists in the electrochemical reduction of Fe, Co
or Ni salts within the unixial pores of an alumina or a polycarbonate
membrane to confine the metal growth \cite{3_Sellmyer_2001}. In order
to grow NWs with a diameter below 10 nm, other solid hosts have been
employed such as carbon nanotubes \cite{4_Tilmaciu_2009} or mesoporous
silica \cite{5_Campbell_2006}. Recently, cobalt NWs with diameters
below 5 nm embedded in epitaxial CeO$_{2}$ layers were obtained by
pulsed-laser deposition \cite{6_ Vidal_2009}. Iron-boride NWs with
diameter in the range 5-50 nm were obtained by chemical vapour deposition
\cite{7_Li_2006}. Wet chemistry methods, for which the adsorption
of long chain molecules at the metal particle surface is the driving
force of the anisotropic growth, were developed for the synthesis
of cobalt \cite{8_Dumestre_2002} and cobalt-nickel NRs and NWs \cite{9_Soumare_2008}.\\

One interest of high aspect ratio ferromagnetic nanoparticles is that
they can present high coercivity properties due to their large shape
anisotropy. Magnetic NWs have been recently proposed as potential
candidates for high-density magnetic recording \cite{10_Xie2012,11_Pak2011},
nanowire-based motors \cite{11_Pak2011} and the bottom-up fabrication
of permanent magnets \cite{10_Maurer_2007}. Indeed, very high coercivity
values were observed on assemblies of cobalt NRs or NWs obtained using
wet chemistry, either by the polyol process \cite{11_Soumare} or
by organometallic chemistry \cite{12_Soulantica}. Composite materials
of well aligned cobalt rods with a volume fraction of 50\% could exhibit
higher (BH)max. values than AlNiCo or ferrite-based magnets \cite{11_Pak2011}.
The interest of these liquid phase processes for the synthesis of
NRs with hard magnetic properties lies in the possibility to produce
wires combining very good crystallinity and small diameter. The cobalt
NRs and NWs synthesized by these chemical methods crystallize with
a hcp structure with the c axis along the long axis \cite{8_Dumestre_2002,13_Viau}.
The coincidence of the shape anisotropy easy axis and the magneto-crystalline
anisotropy easy axis reinforces the whole magnetic anisotropy \cite{10_Maurer_2007,12_Soulantica}.

Several parameters govern the coercivity of an assembly of anisotropic
magnetic particles: (i) the distribution of the particle easy axis
orientation with respect to the magnetic field and (ii) the particles
shapes. For cobalt wires we showed that the coercivity increases when
they are aligned parallel to the applied magnetic field in good agreement
with the Stoner-Wohlfarth model \cite{11_Soumare}. The coercivity
is the sum of the magnetocrystalline and the anisotropy contributions.
The contribution due to the shape anisotropy actually strongly depends
on the detailed geometry of the wires and especially of the wire tips.
This was investigated by micro-magnetic simulations \cite{14_Ott}.
On the one hand, while the ellipsoid shape is the most favourable,
a cylindrical shape also provide a good shape anisotropy. On the other
hand, enlarged tips create nucleation points for the magnetization
reversal which can significantly reduce the coercivity by up to $30
$. Surface effects related to the thin superficial CoO layer were extensively
studied \cite{16_Maurer}. The measured Néel temperature of the CoO
shell was 230 K. Significant modifications of the magnetic behaviour
take place below this temperature. When the temperature decreases,
a coercivity drop is actually observed below 150 K \cite{16_Maurer}.\\

But even if progresses have been made in the synthesis of ferromagnetic
NRs and NWs exhibiting large coercivities at room temperature and
in the understanding of their magnetic properties \cite{10_Maurer_2007,11_Soumare,14_Ott,15_Maurer,16_Maurer},
there is still a lack of information about the stability of both their
structural and magnetic properties above room temperature which is
a key information for any practical use at high temperatures of these
materials e.g. for the fabrication of rare-earth free permanent magnets.
At high temperatures, metal wires generally undergo an irreversible
transformation to chains of spheres, described as the Rayleigh instability\cite{17_Karim,18_Nisoli}.
The high aspect ratio of compacted wires may also be altered by sintering.
A first study showed that the thermal stability of \textquotedblleft{}organometallic\textquotedblright{}
cobalt nanowires was dependent on the atmosphere under which the wires
were annealed \cite{19_Ciuculescu}. Fragmentation of cobalt NWs into
chains of cobalt particles due to the Rayleigh instability was avoided
when the cobalt NWs were coated by a thin carbon shell \cite{19_Ciuculescu}.\\

In this paper we present the high temperature structural and magnetic
properties of cobalt nanorods prepared by the polyol process using
in-situ characterizations. The scope of this communication is to determine
whether the anisotropic structure and texture are modified at high
temperature and consequently whether their magnetic properties, and
especially their large coercivities, are preserved. The temperature
range of stability of these NRs and the temperature dependence of
their intrinsic magnetic properties up to 623 K under different atmospheres
are described.

\section{Materials and methods.}

CoCl$_{2}\cdot$6H$_{2}$O (Alfa Aesar, 99.9$\%$), RuCl$_{3}\cdot$xH$_{2}$O
(Aldrich, 99.98$\%$), NaOH (Acros), 1,2- butanediol (Fluka, $\geq98\%$),
methanol (VWR, Normapur) and sodium laurate, Na(C$_{11}$H$_{23}$COO)
(Acros, 98$\%$) were used without any further purification.

\subsection{Syntheses of Co nanowires. }

The cobalt laurate precursor, Co$^{\textrm{II}}$(C$_{11}$H$_{23}$COO)$_{2}$
was first prepared. In 100 mL of distilled water at 333 K, was dissolved
sodium laurate (75 mmol; 16.7 g) and to this solution, was added an
aqueous solution (38 mL) of cobalt(II) chloride (40 mmol, 10,0 g)
pre-heated at 333K under vigorous stirring. This resulted in the formation
of a purple precipitate which was vigorously stirred at 333K for 15
min. The Co(II) solid phase was washed twice with distilled water
(100 mL) then with methanol (100 mL) and finally dried in an oven
at 323 K overnight. Yield: 94$\%$ (based on Co).

The synthesis of the cobalt NRs was realized according to a procedure
previously described \cite{11_Soumare}. To 75.0 mL of 1,2-butanediol
were added Co(C$_{11}$H$_{23}$COO)$_{2}$ (2.75 g, 0.08 M), RuCl$_{3}$.xH$_{2}$O
(3.2 10$^{-2}$ g) and NaOH (0.225 g, 0.075 M). The mixture was heated
to 448 K with a ramping rate of 13 K.min$^{-1}$ for 20 min until
the color of the solution turned black, indicating the reduction of
Co(II) into metallic cobalt. After cooling to room temperature, the
Co NRs were recovered by centrifugation at 8500 r.p.m. for 15 min,
washed with 50 mL of absolute ethanol (3 times), and finally dried
in an oven at 323 K. Yield: $92\%$ (based on Co). Elemental analyses
revealed C and H amounts of 3.8 and 0.7 wt. $\%$ , respectively.

\subsection{Preparation of the sample for magnetic measurements.}

Two sets of samples were prepared for magnetic measurements. Sample
(A): a few drops of a Co NRs suspension in toluene were deposited
on an aluminum foil and the toluene was removed by evaporation under
the application of an external magnetic field of 1 T. Sample (B):
a pellet of magnetic NRs was prepared using an infrared KBr die (internal
diameter 13 mm) and applying a pressure of about 6 tons delivered
by a hydraulic press. The mass and the apparent density (including
porosity) of the pellet were respectively 0.26 g and 2.13 g cm$^{\lyxmathsym{\textendash}3}$.
The true density of the powder, obtained at 298 K using a helium Accupyc
1330 pycnometer from Micromeritics, was found to be $6.76\pm0.41$
g cm$^{\lyxmathsym{\textendash}3}$. The packing factor, defined as
the ratio of the volume of the particles by the volume of the pellet
was about $30\%$. Co nanowires are randomly oriented inside the pellet
since no magnetic field was applied during its preparation. The small
density of the powder is accounted for by the fact that part of the
metallic wires are oxidized and that some organic material remain
in the pellets.

\subsection{Characterization techniques. }

\subsubsection{Room temperature characterizations}

Transmission electron microscopy (TEM) characterizations were performed
using a Jeol 100-CX II microscope operating at 100 kV. Infrared spectra
were recorded on a nitrogen purged Nicolet 6700 FT-IR spectrometer
equipped with a VariGATR accessory (Harrick Scientific Products Inc.,
NY) fixing the incident angle at $62^{\circ}$. A drop of the colloidal
solution was deposited on the Ge wafer and the spectrum was recorded
when the solvent (absolute ethanol or toluene) was fully evaporated.
XRD patterns obtained using Co K radiation ($\lambda=1.7889$ $\AA$)
were recorded on a PANalytical X\textquoteright{}Pert Pro diffractometer
equipped with an X\textquoteright{}celerator detector in the range
$20-80{}^{\circ}$ with a $0.067{}^{\circ}$ step size and 150 s per
step. The size of coherent diffraction domains, $L_{hkl}$, were determined
using MAUD software which is based on the Rietveld method combined
with Fourier analysis, well adapted for broadened diffraction peaks.
Magnetic measurements were performed using a Quantum Design MPMS-5S
SQUID magnetometer.

\subsubsection{Thermal treatments and high temperature characterizations.}

In order to follow the structural evolution of the cobalt nanowires
with temperature, insitu thermal treatments were realized in the range
300-673 K, using a HTK 1200N high-temperature X-ray diffraction chamber
from Anton Paar. Two types of experiments were performed, depending
on the oxygen content of the nitrogen gas used: in the first one,
the powder was heated under a high-purity nitrogen atmosphere(O$_{2}$
< 2 ppm; N$_{2}$ Alphagaz 1) while in the second one, the oxygen
content was substantially higher (O$_{2}$ < 0.1 ppm; N$_{2}$ Alphagaz
2). The samples were heated inside the high temperature X-ray diffraction
chamber from room temperature to the final temperature using a 5 K·min$^{-1}$
rate and maintained for 2 h at the final temperature before the acquisition
of a X-ray diffraction pattern. Magnetization curves at high temperature
of samples (A) and (B) were measured with a SQUID equipped with an
oven. In this procedure the particles were heated in a reduced helium
pressure.

\section{Results and discussions}

\subsection{Morphology and chemical analysis of the cobalt nanorods.}

TEM observations on the particles prepared by the polyol process showed
Co NRs with a mean diameter$d_{m}=13$ nm and a mean length $L_{m}=130$
nm (Figure 1a). The standard deviation of the diameter is very small($<15\%$
of the mean length) as it was observed before \cite{11_Soumare}.
For the Co NRs deposited on an Al substrate, the application of an
external magnetic field results in the alignment of the cobalt anisotropic
nanoparticles while the Co NRs are randomly oriented within the pellet
(see Figure 1b and 1c). Moreover Figure 1c shows also that the compression
was not detrimental to the Co NRs. The Co NRs were further characterized
using TG-DT analyses under pure N$_{2}$ (O$_{2}$ < 0.1 ppm). The
thermogram (see Figure 2) showed a $7\%$ weight loss at about 575
K associated with a sharp endothermic peak. This weight loss is attributed
to the elimination of organic matter adsorbed on the particle surface.
Infrared spectroscopy was thus performed on the cobalt particles to
characterize the organic matter remaining after the synthesis and
the washing procedure at room temperature. The cobalt NRs infrared
spectrum washed twice with ethanol (Figure 3) exhibits a large band
centered at 3300 cm$^{-1}$ corresponding to the O-H stretching vibration.
The OH groups can belong to a surface cobalt hydroxide and/or to adsorbed
ethanol. At 2850 and 2920 cm$^{-1}$ the symmetric and asymmetric
C-H stretching vibration are respectively observed, which is attributed
to the CH$_{2}$ groups of the laurate ions. In the region between
1400 and 1550 cm$^{-1}$ the intense bands are attributed to the asymmetric
and symmetric C-O stretching vibration of carboxylate groups, indicating
that laurate ions remain at the particle surface. The intensity of
all these bands decreases with the successive washings (Figure 3)
showing that the amount of organic ligands at the particle surface
is strongly dependent on the way they have been washed.

\begin{figure}
\includegraphics[bb=70bp 150bp 600bp 585bp,clip,width=8.5cm]{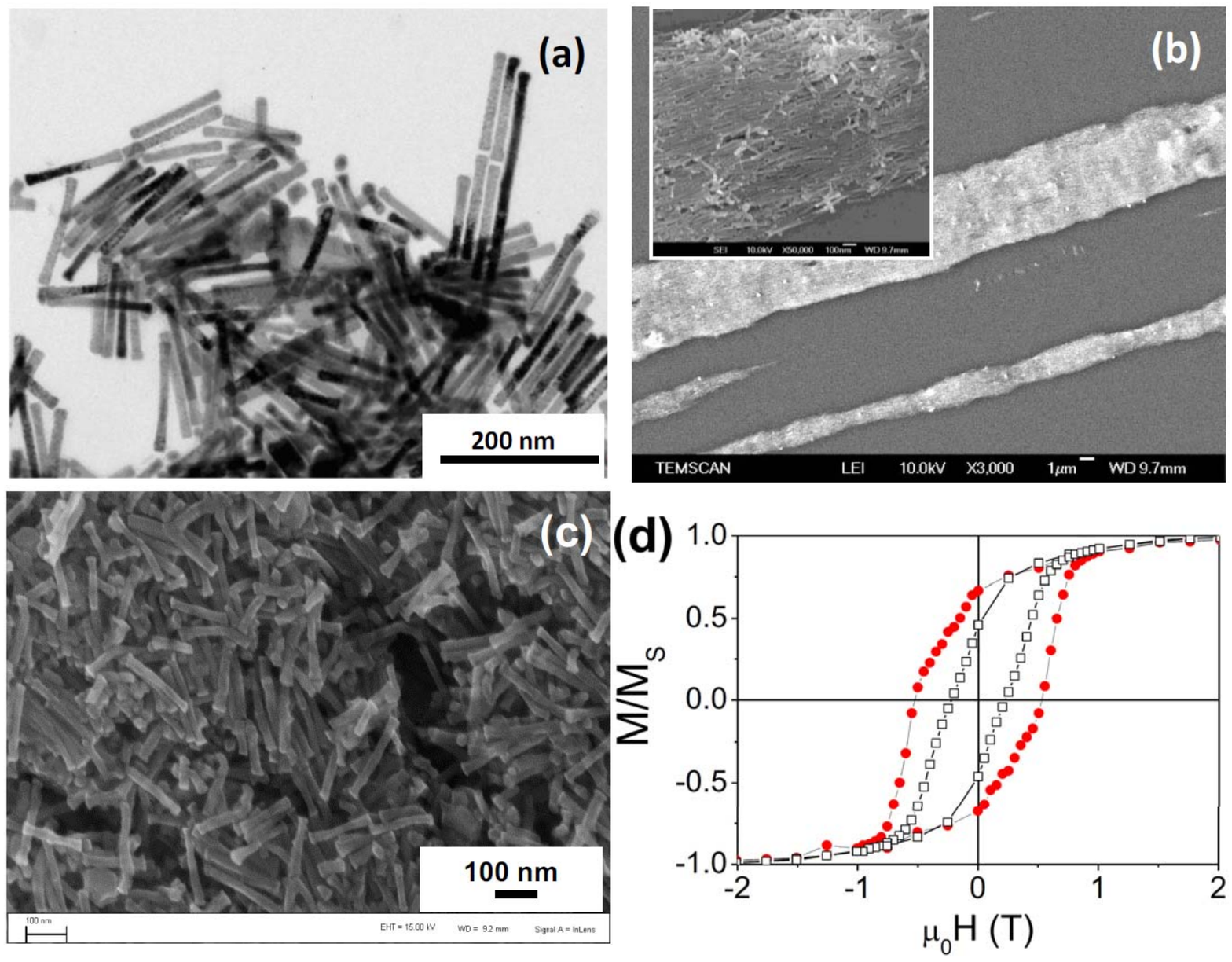}

\caption{(a) TEM images of Co rods ($L_{m}=130$ nm, $d_{m}=13$ nm); SEM images
of (b) sample (A), (c) sample (B) and (d) magnetization curves at
300 K of sample (A) (circles) and (B) (squares).}
\end{figure}

\begin{figure}
\includegraphics[bb=30bp 150bp 600bp 595bp,clip,width=8cm]{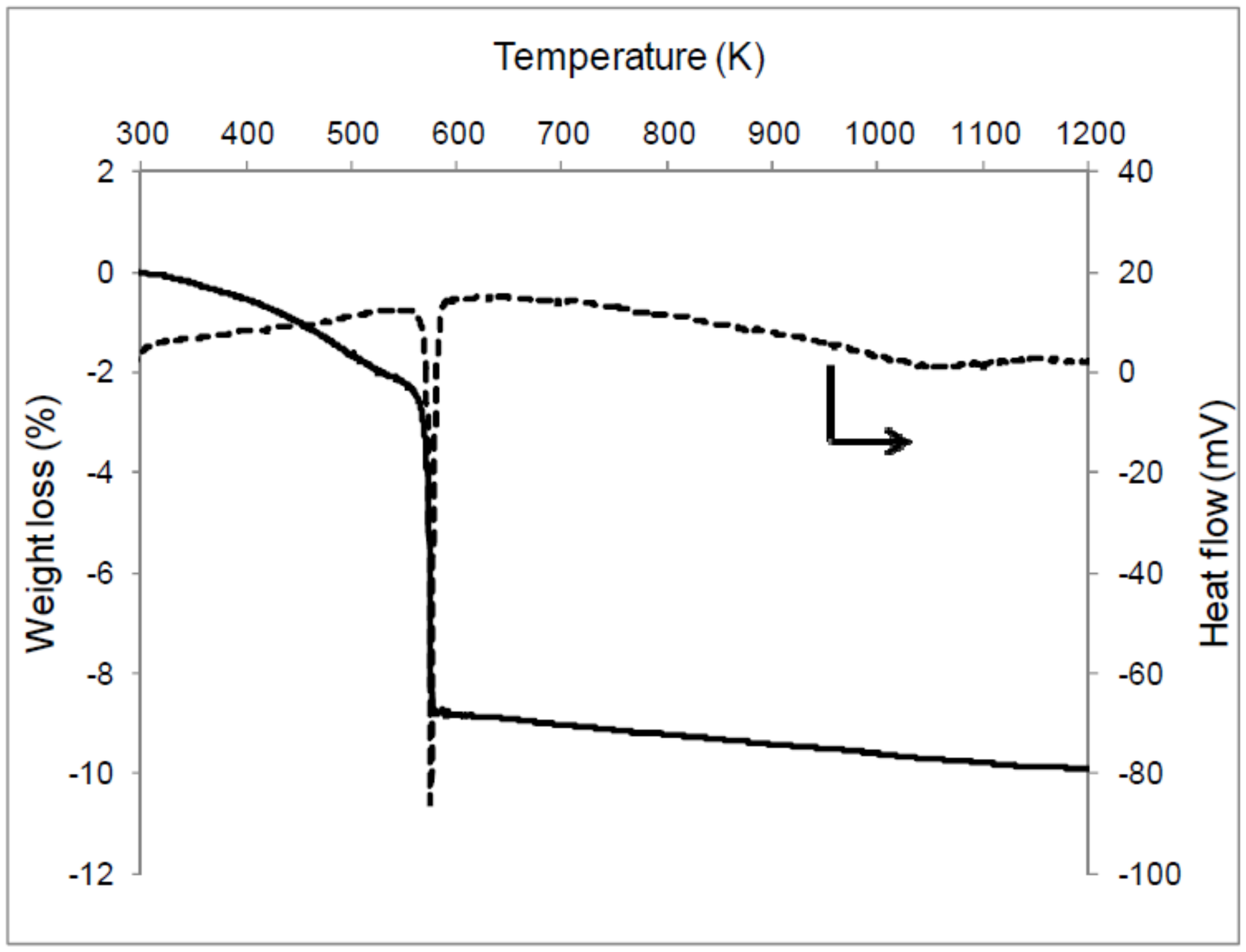}

\caption{Thermogravimetric and differential thermal analyses realized on Co
nanowires under N$_{2}$ (O$_{2}<$ 0.1 ppm).}
\end{figure}

\begin{figure}
\includegraphics[bb=30bp 100bp 630bp 595bp,clip,width=8cm]{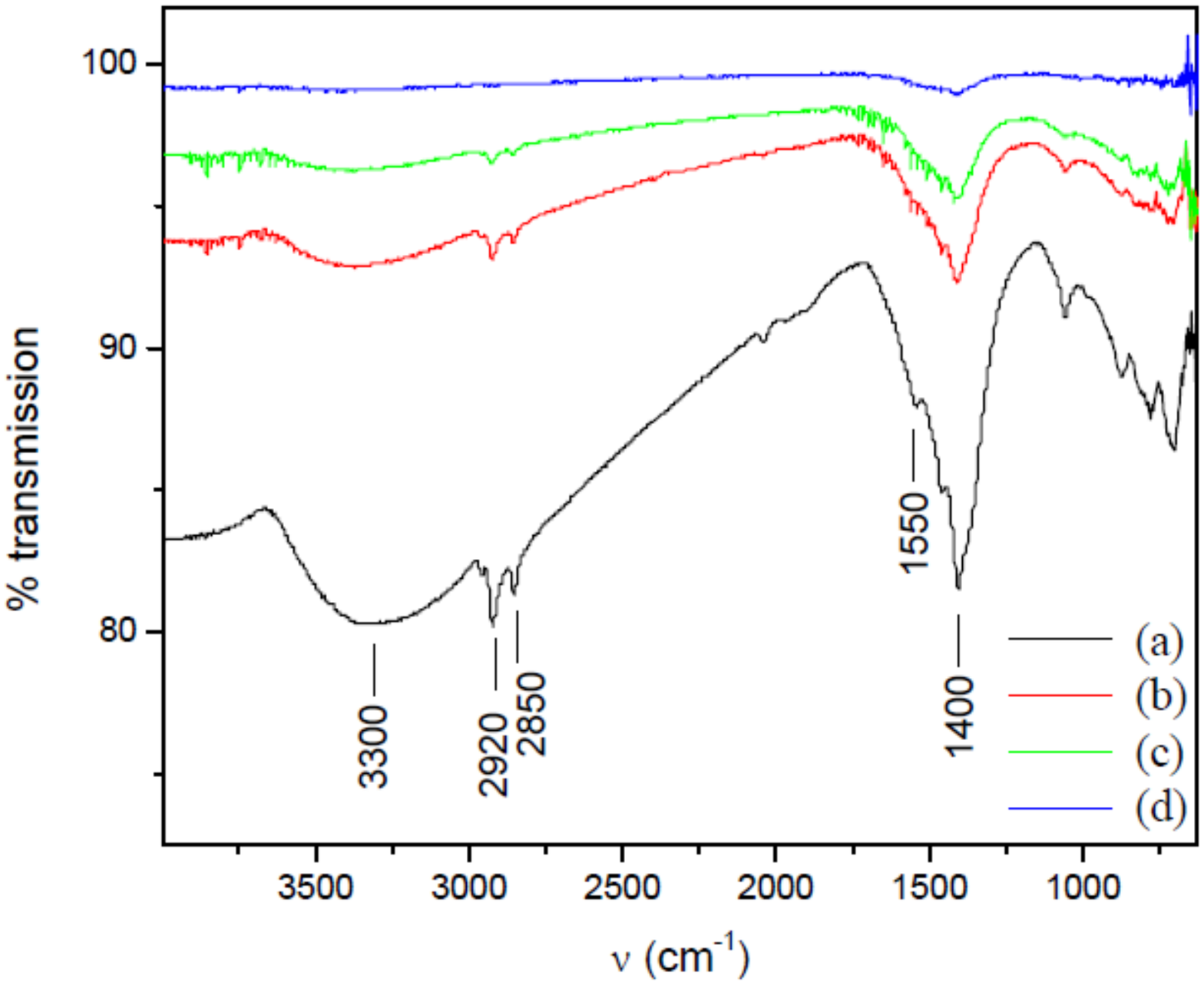}

\caption{Infrared absorption spectra of cobalt wires washed twice with ethanol
(a); washed twice with ethanol and once (b), twice (c) and three times
(d) with toluene.}
\end{figure}

\subsection{High temperature structural and chemical modification. }

In order to follow the structural modifications of the Co NRs when
they are heated up, in-situ X-ray diffraction have been performed
for different temperatures ranging from 298 K up to 623 K. The structural
and chemical modifications were found to depend on the atmosphere
under which the cobalt wires are annealed. The in-situ X-ray diffraction
patterns of the Co NRs annealed under a N$_{2}$ atmosphere with O$_{2}$
concentration $<$ 2 ppm at different temperatures showed a progressive
oxidation emphasized by the growth of the CoO (111) peak around $42{}^{\circ}$in
spite of the small O$_{2}$ concentration. With increasing temperature,
the very broad peaks of the oxide becomes thinner indicating grain
growth processes and/or crystallization of a pre-existing amorphous
phase. Previous HRTEM studies performed on Co nanowires have evidenced
the presence of a CoO layer composed of disoriented crystallites of
various sizes \cite{16_Maurer}, indicating that the former case is
the more probable one. No structural modification was observed on
the hcp non oxidized cobalt phase. For Co samples treated under N$_{2}$
with very low dioxygen content ($<$ 0.1 ppm) the Xray diffraction
data (Fig. 4) show that there is also development of the CoO oxide
between 298K and 473 K. Then, between 473 and 523 K, CoO vanished
at the benefit of metallic Co as indicated by the sharpening and the
increase of intensity of the Co peaks. For higher temperatures, 573
K and 623 K, CoO is detected again. Given the high temperature applied,
the very low dioxygen content is nevertheless sufficient to induce
the formation of this oxide. For the as-synthesized sample, TG-DT
analyses realized under N$_{2}$ (O$_{2}$ content < 0.1 ppm) have
shown a weight loss of about 7 \% at 573 K associated with an endothermic
signal (see Figure 2). This weight loss is explained by the decomposition
of the remaining metal-organic species adsorbed at the particle surface.
Indeed, IR-ATR experiments (Figure 3) have clearly shown vibrations
attributed to adsorbed laurate species (see above). Thus, the comparison
of the HT XRD patterns with the IR and TGA results suggest that the
organic molecules remaining at the particle surface reduce the CoO
shell in the temperature range corresponding to their decomposition.
This effect can be evidenced only when the oxygen concentration in
the atmosphere is small enough.

\begin{figure}
\includegraphics[bb=30bp 35bp 430bp 580bp,clip,width=8cm]{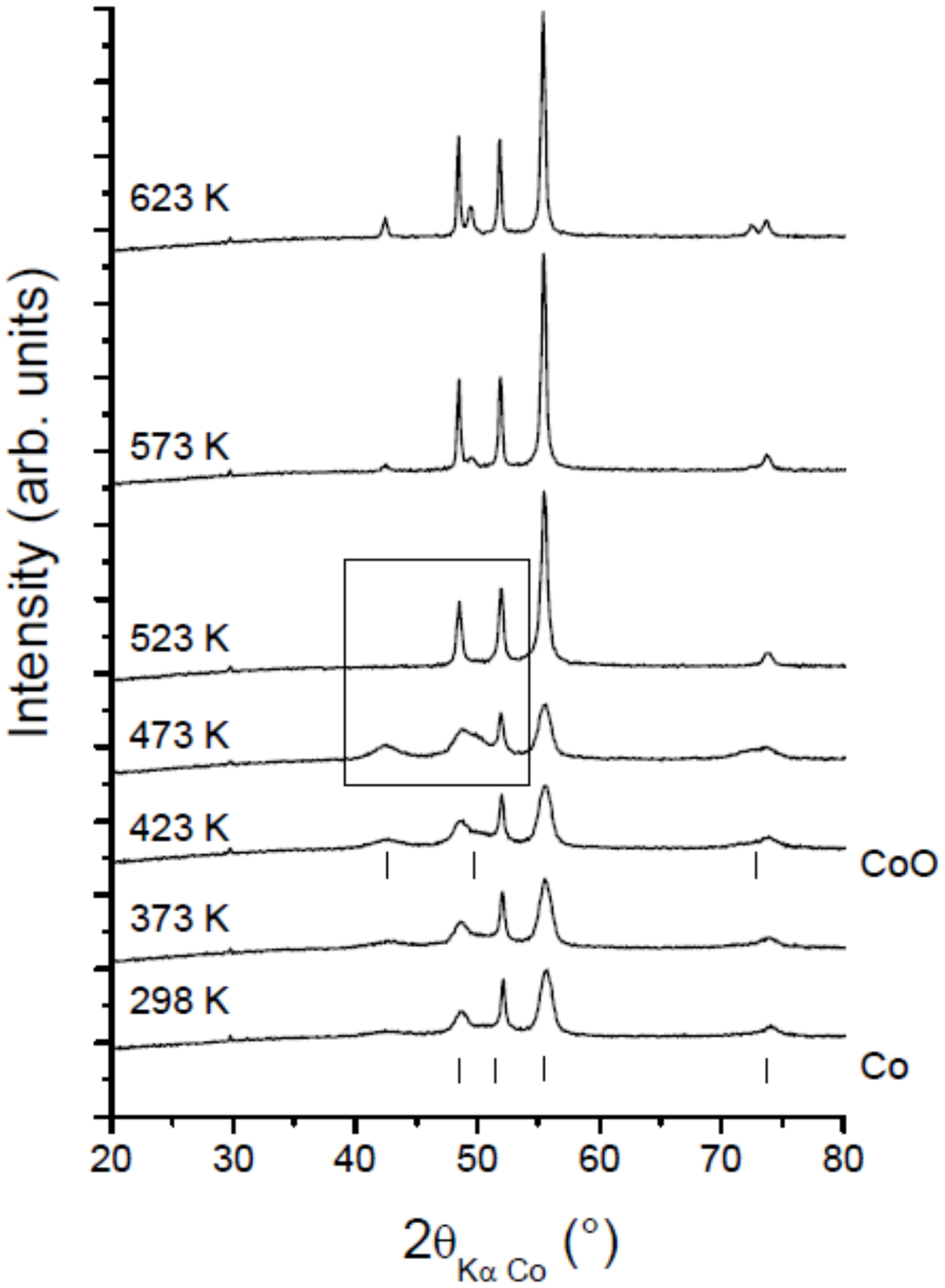}

\caption{In-situ X-ray diffraction patterns of Co nanorods thermally
treated under N$_{2}$ (O$_{2}<$ 0.1 ppm).}
\end{figure}

\subsection{High temperature texture modification.}

The XRD diffractograms showed that the hexagonal close-packed structure
of the Co core is preserved for temperatures up to 623 K, whatever
the annealing atmosphere. Nevertheless, a close examination of these
patterns shows that the line broadening is modified during the thermal
treatment. The mean crystallite sizes for the (10.0) and (00.2) reflexions
have been determined and plotted as a function of the temperature
treatment for samples annealed in N2 with the lowest oxygen content
(Figure 5). At room temperature, the mean crystallite size for the
(10.0) reflexion is always found much smaller than the mean crystallite
size for the (00.2) reflexion. This observation confirms that the
long axis of the cobalt wires is the c axis of the hcp structure.
The data show that up to 500 K, the (10.0) and (00.2) mean crystallite
sizes are more or less constant, indicating that the anisotropic shape
of the crystallites is preserved. Major texture modifications are
observed above 525K. Indeed, at this temperature a considerable increasing
of both the L$_{10.0}$ and L$_{00.2}$ mean crystallite sizes is
observed and the crystallites lose their anisotropy. At 525 K particles
start to undergo sintering. In order to probe a possible deterioration
of the Co nanowire morphology, TEM was performed ex-situ for samples
annealed at different temperatures. Figures 6a and 6b indicate that
the shape of the nanowires is kept up to 523 K. On the other hand,
Figure 6c shows that at 573 K and above, the nanowires start to sinter
so that their shape anisotropy starts to fade away at these temperatures.
At 623 K, the anisotropic shape is lost and only large aggregates
are observed (Figure 6d). The sintering of the particles is probably
enhanced by the fact that the CoO layer which likely acts as a protective
layer preventing sintering is reduced to cobalt by the organic molecules
at about 523 K as seen by XRD (Figure 4).

\begin{figure}
\includegraphics[bb=30bp 20bp 760bp 580bp,clip,width=8cm]{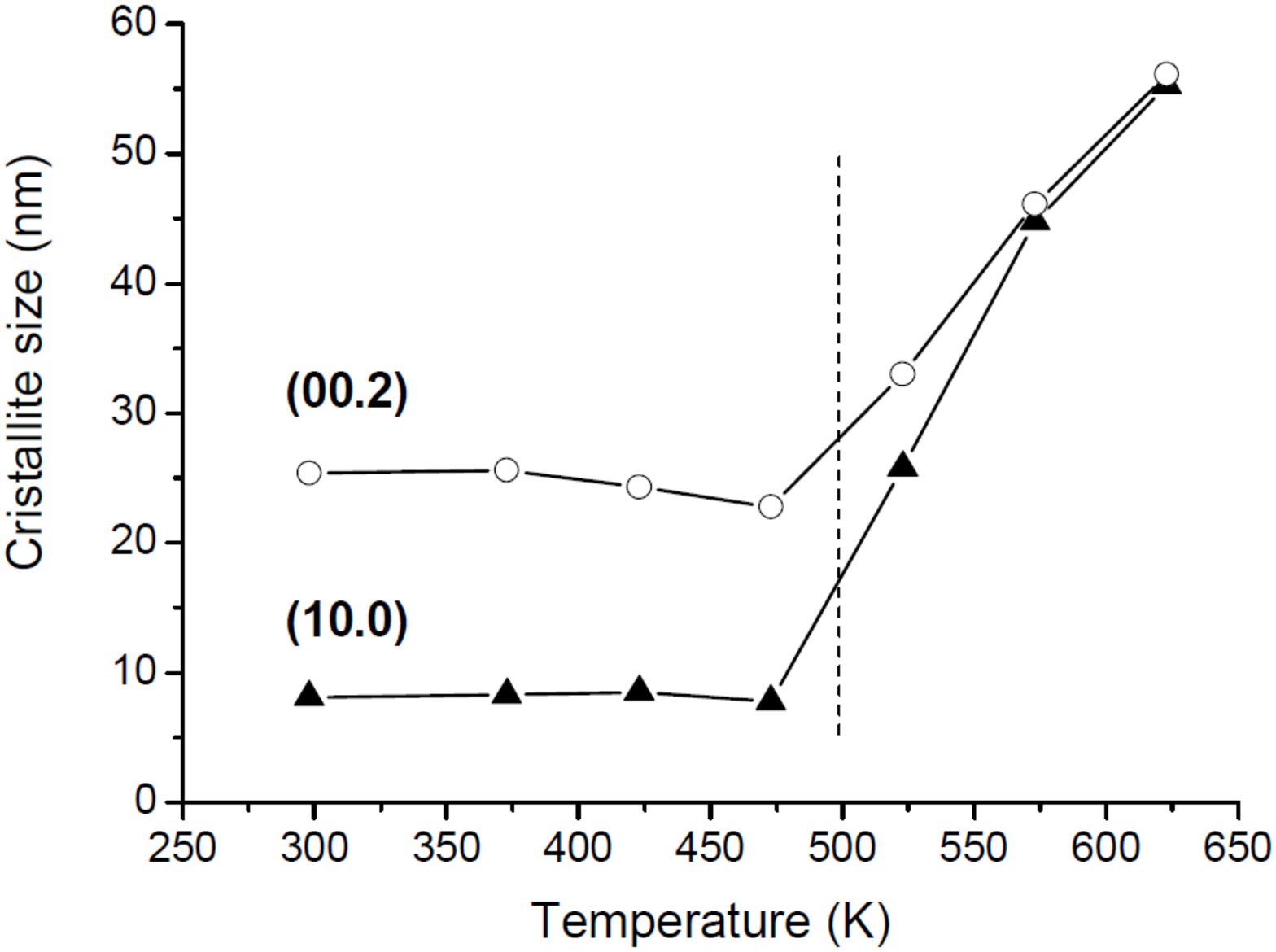}

\caption{Variation of the L$_{10.0}$ and L$_{00.2}$ mean crystallite
sizes with increasing temperature for Co samples treated under
N$_{2}$ (O$_{2}<$ 0.1 ppm).}
\end{figure}

\begin{figure}
\includegraphics[bb=30bp 140bp 480bp 595bp,clip,width=8cm]{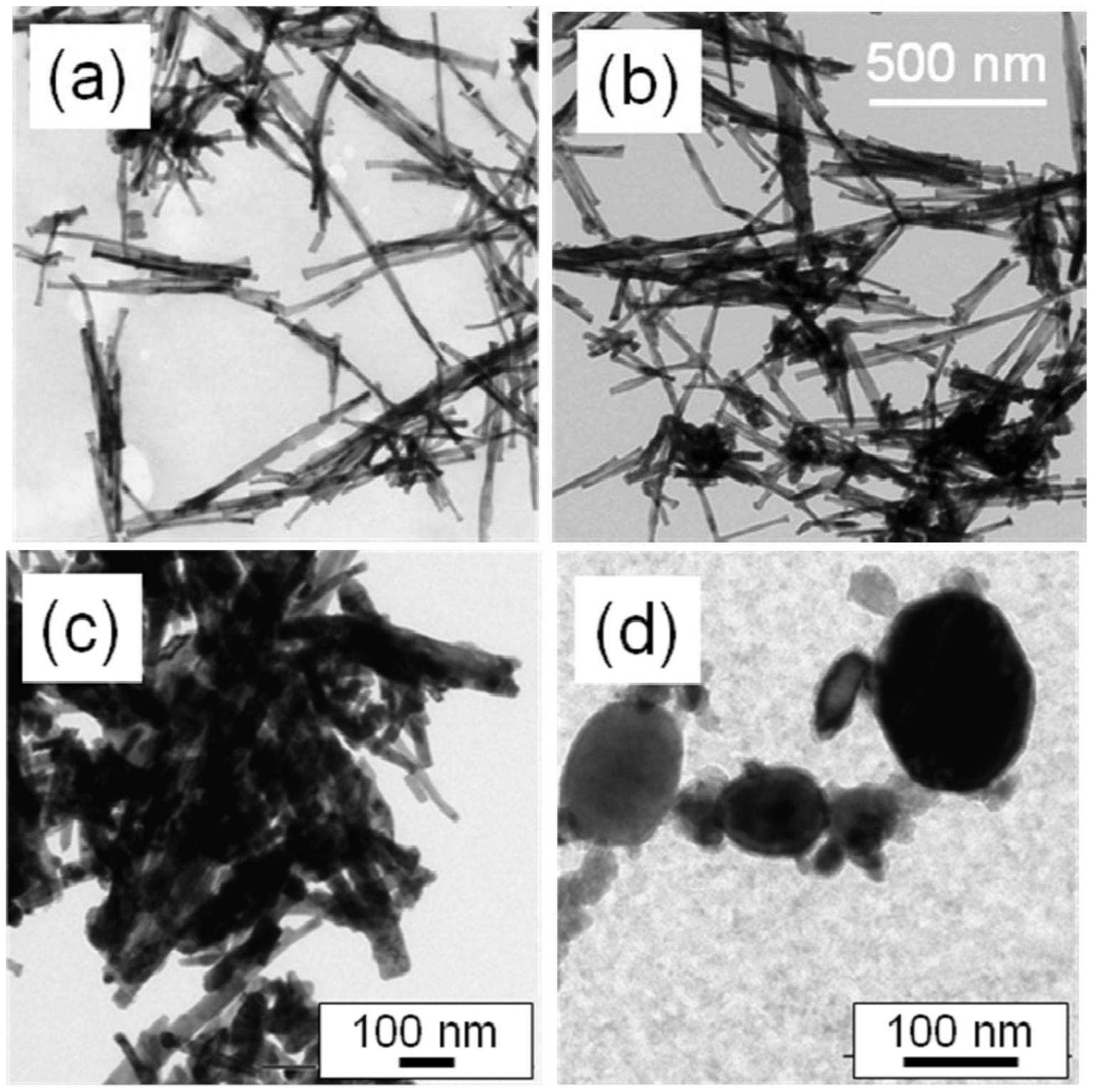}

\caption{TEM images of Co wires (a) after drying at 330 K and after
thermal treatment under N$_{2}$ (O$_{2}$ < 0.1 ppm) for 2 hours at
523 K (b), at 573 K (c) and 2 hours at 623 K (d).}
\end{figure}

\subsection{Magnetic properties at high temperature. }

The cobalt NRs are ferromagnetic at room temperature with coercivities
of 530 and 230 mT and $M_{r}/M_{S}$ values of 0.67 and 0.46 for the
samples (A) and (B), respectively. The difference of $H_{C}$ and
$M_{r}$ are essentially due to the high degree of orientation of
the rods in the sample (A) which increases both coercivity and remanence.
The saturation magnetization $M_{S}$ of sample (B) is 113 emu.g$^{-1}$
(1.410$^{\lyxmathsym{\textendash}4}$ T.m$^{3}$.kg$^{-1}$). Such
a low value (compared to the bulk value of 165 emu.g$^{-1}$) is due
to the surface oxidation of the wires and by the remaining organic
matter in the pellet. The density of the pellet, determined using
He pycnometry, was found to be 6.76$\pm$0.41 g.cm$^{-3}$. Taking
into account that the organic matter in the particles corresponds
to about 7 wt. $\%$ and that the CoO layer thickness is 1.2 nm for
a nanorod with a mean diameter of 15 nm and a mean length of 130 nm
\cite{16_Maurer}, a ratio of 73 wt. $\%$ of metallic Co to the total
mass of the solid (including CoO and the organic matter) can be estimated.
This value is in good agreement with that determined using the apparent
$M_{S}$ value (113/165 = 68$\%$ ). The temperature dependence of
the saturation magnetization is presented on Figure 7a. Two very different
behaviours are observed. In the case of Sample (B) (pressed pellet),
the saturation magnetization $M_{S}$ is rather stable up to 550 K;
$M_{S}$ decreases by less than 1$\%$ from 300 K to 550 K. At 550
K, a jump of the magnetization is observed corresponding to an increase
of 25$\%$ with respect to the room temperature magnetization value.
At higher temperatures (up to 800 K), the saturation magnetization
decreases only moderately (10$\%$). The magnetization jump at 550
K is irreversible and can be explained by the decomposition of remaining
metal-organic at the surface of the NRs that reduces the cobalt oxide
layer and causes the apparition of added metallic Co. This phenomenon
was clearly evidenced by XRD (Figure 4). For the sample (A), the saturation
magnetization of the cobalt NRs deposited on an aluminium foil varies
only slightly up to 500 K but decreases strongly above this temperature.
In this case, the decrease of the magnetization is explained by the
oxidation of the cobalt NRs into CoO as was inferred by XRD. The successive
washings have removed most of the organic matter (Figure 3) and no
reduction can occur. The helium reduced pressure in the oven of the
SQUID was not sufficient to prevent the powders from oxidation at
high temperature since traces of O$_{2}$ were probably present.

The temperature dependence of the coercivities are presented in Figure
7b. When normalized to the to the room temperature coercivity, $H_{C}(T)/H_{C}(300K)$,
the coercivity of both samples follows the same behaviour over the
temperature range 300-500K. The coercivity $\mu_{0}H_{C}$ follows
a linear dependence:

\[
\frac{\mu_{0}H_{C}}{\mu_{0}H_{C}(300K)}=1\lyxmathsym{\textendash}a(T\lyxmathsym{\textendash}300)
\]

where $a$ is 2.4$\times$10$^{\lyxmathsym{\textendash}3}$ K$^{\lyxmathsym{\textendash}1}$.
This dependence can be accounted for by the temperature dependence
of the magneto-crystalline anisotropy of hcp cobalt. The magneto-crystalline
anisotropy of mono-crystalline cobalt is usually described by an anisotropy
energy of the form $Ea=K_{u1}\sin^{2}\theta+K_{u2}\sin^{4}\theta$
with $K_{u1}=4.1\times10^{5}$ J.m$^{-3}$ and $K_{u2}=1.4\times10^{5}$
J.m$^{-3}$ at 300 K and is $\theta$ the angle of the magnetization
with respect to the c axis. The respective temperature variations
of these anisotropy constants is however non-trivial: while the value
of $K_{u2}$ monotonously decreases down to $0.4\times10^{5}$ J.m$^{-3}$
at 600K, the value of $K_{u1}$ changes sign at 520 K \cite{22_Ono }.
The spin reorientation (from parallel to perpendicular to the c-axis)
is mainly driven by the change of sign of $K_{u1}$ at 520 K, the
effect of $K_{u2}$ being to \textquotedblleft{}smear\textquotedblright{}
the transition so that the reorientation from $0{}^{\circ}$ to $90{}^{\circ}$takes
place over a rather wide temperature range (520-600 K). At 550 K,
the spin reorientation is about $40{}^{\circ}$ \cite{23_Barnier1961}
with a magneto-crystalline energy as low as $0.1\times10^{5}$ J.m$^{-3}$,
that is less than a 1/50 of the room temperature anisotropy so that
the Co NR can be considered to have no more magneto-crystalline anisotropy.

In the case of sample B, sintering of the wires interferes with the
intrinsic magnetic properties of the Co NRs. On the other hand, in
the case of sample A, the wires are structurally stable in shape at
least up to 600 K and the wires are rather well aligned with a dispersion
in their directions smaller than $7{}^{\circ}$ \cite{15_Maurer}.
We can thus consider sample A as consisting of individual particles
behaving as Stoner-Wohlfarth particles. At 550K when the magneto-crystalline
anisotropy vanishes, the coercivity of elongated ellipsoidal particles
with their long axis aligned with the applied field can be expressed
as $\mu_{0}H_{C}=2K_{shape}/M_{S}$ \cite{24_Handley} where $K_{shape}$
corresponds to the effective anisotropy constant related to the shape
anisotropy of the particles. Note that this formula is strictly valid
only if the magnetization rotation is coherent, which requires that
the particle diameter is very small ($\approx$ 8 nm for Co). Our
NRs dimensions are close to this limit. The value of the coercivity
at 550K is $\mu_{0}H_{C}\approx210$ mT which thus corresponds to
the contribution of the shape anisotropy. This represents only half
of the theoretical value for an aspect ratio of 5 ($2K_{shape}/M_{S}\approx527$
mT) . The fact that the measured value is significantly smaller than
the maximum theoretical value for equivalent ellipsoid can be easily
accounted for by the fact that: (i) the NRs assembly is not well aligned
which dramatically reduces the coercivity related to the shape anisotropy,
(ii) for an equivalent aspect ratio, ellipsoids have a higher coercivity
compared to cylinders because of domain nucleation \cite{14_Ott}
and (iii) we are not working at 0 K so that thermally activated reversal
plays a significant role.

\begin{figure}
\includegraphics[bb=30bp 80bp 380bp 595bp,clip,width=8cm]{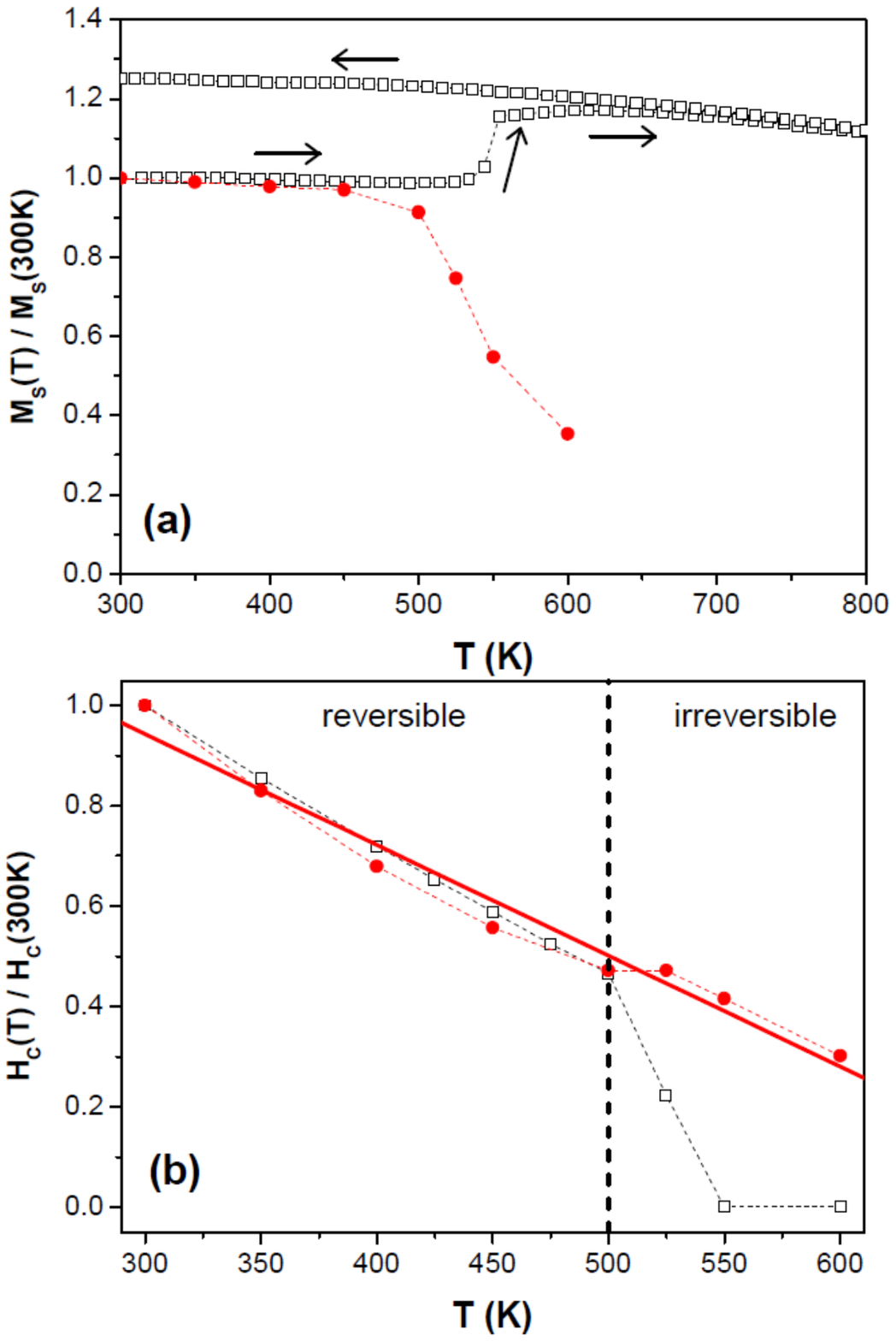}

\caption{$M_{S}(T)/M_{S}(300K)$(a) and $H_{C}(T)/H_{C}(300K)$ (b) as a function
of temperature as a function of temperature for sample (A) and sample
(B). The solid line is a linear fit to the experimental data corresponding
to sample (A), which is given by $\frac{\mu_{0}H_{C}}{\mu_{0}H_{C}(300K)}=1\lyxmathsym{\textendash}a(T\lyxmathsym{\textendash}300)$
(see text for details).}
\end{figure}

From a point of view of using Co NRs as a basis for the fabrication
of permanent magnetic materials, one important aspect was to evaluate
magnetic dipolar magnetic interactions between nanowires in dense
magnetic aggregates. This has been very recently addressed and micromagnetic
simulations have shown that dipolar interaction between wires are
not detrimental to the high coercivity properties, even for very dense
aggregates \cite{15_Maurer}. With respect to high temperature operation,
the present study shows that the temperature dependence of the Co
magneto-crystalline anisotropy is a real limitation. $K_{MC}$ not
only vanishes around 550 K but becomes negative above this temperature,
which, as a consequence, leads to magnetic properties which are very
sensitive to the temperature. Nevertheless, the materials are rather
resilient to high temperatures (up to 550 K) and are competitive with
the other existing types of permanent magnetic materials. The magnetic
performances of Co nanowires are of the order of$(BH){}_{max}\approx$
12-15 MG Oe at room temperature \cite{10_Maurer_2007} which rank
them in-between ferrites ($\approx$ 3 MG Oe) and AlNiCo ($\approx$
5 MG Oe) and RE-magnets (NdFeB $\approx$ 40 MG Oe; SmCo $\approx$
20-30 MG Oe). A general drawback of rare earth based materials is
their relatively lower Curie temperature (580 K for NdFeB and 1023
K for SmCo) which leads to a significant dependence of the magnetization
at high temperatures which is not the case for pure Co systems ($T_{C}\approx$1300
K). NdFeB magnets cannot be operated at temperatures above 400 K because
of irreversible losses. SmCo magnets can be operated at elevated temperatures
(up to 523 K typically) which is equivalent to our materials. They
have a similar temperature coefficient for the coercive field ($\approx-2\times10^{-3}$K$^{\lyxmathsym{\textendash}1}$)
as Co nanowires.

\section{Conclusion}

The interest of the Co NRs prepared by the polyol process lies in
their well-controlled morphology, diameter smaller than 15 nm, high
aspect ratio and shape homogeneity, and, consequently, in their high
coercivity at room temperature. In this paper, we have presented a
study on structural and magnetic properties of these particles at
high temperatures. In the temperature range 300-550 K, we show that
the coercivity decreases linearly and that this variation is reversible.
This can be accounted for by the temperature dependence of the magneto-crystalline
anisotropy of cobalt. At 550 K, 40$\%$ of the room temperature coercivity
is maintained. This value corresponds to the contribution of the shape
anisotropy to the global anisotropy since at 550K the magnetocrystalline
contribution has vanished. Above 525 K, the magnetic properties are
irreversibly altered either by sintering or by oxidation. In absence
of oxygen the decomposition of metal-organic matter remaining at the
particle surface reduces the native cobalt oxide layer and provokes
coalescence. The coercivity is irreversibly altered by the loss of
shape anisotropy and by the particle growth that may induce a transition
from mono- to multi- domains magnetic particles. In slightly oxidative
conditions, the growth of a thin oxide layer at the wire surface prevents
from sintering but decreases significantly the saturation magnetization.
In terms of temperature stability some further work is necessary to
prevent both coalescence and oxidation of the wires at high temperatures
(>550K). The use of a passivation layer is a possible route to that.
\begin{acknowledgments}
The authors gratefully acknowledge the Agence Nationale de la Recherche
for their financial support (project 07-NANO-009 MAGAFIL). We thank
F. Herbst (ITODYS) for providing the TEM images of nanowires and J.B.
Moussy (CEA-IRAMIS) for his help in the magnetometry measurements.
The help of Dr. S. Khennache for He pycnometry measurements was greatly
appreciated.\end{acknowledgments}


\begin{thebibliography}{References}
\bibitem{1_lieber_2007} Lieber, C. M.; Wang, Z. L.; MRS Bull. 2007,
32, 99.

\bibitem{2_xia_2003} Xia, Y.N.; Yang, P.D.; Sun, Y.G.; Wu, Y.Y.;
Mayers, B.; Gates, B.; Yin, Y.D.; Kim, F.; Yan, Y. Q. Adv. Mater.
2003, 15, 353.

\bibitem{3_Sellmyer_2001} Sellmyer, D. J.; Zheng, M.; Skomski, R.
J. Phys. : Condens. Matter 2001, 13, R433.

\bibitem{4_Tilmaciu_2009} (a) Tilmaciu, C.M.; Soula, B.; Galibert,
A.M.; Lukanov, P. ; Datas, L.; Gonzalez, J.; Barquin, L.F.; Fernandez,
J.R.; Gonzalez-Jimenez, F.; Jorge, J.; Flahaut, E. Chem. Commun. 2009,
43, 6664; (b) Weissker, U.; Loffler, M.; Wolny, F.; Lutz, M.U.; Scheerbaum,
N.; Klingeler, R.; Gemming, T.; Muhl, T.; Leonhardt, A.; Buchner,
B. J. Appl. Phys. 2009, 106, 054909.

\bibitem{5_Campbell_2006}(a) Campbell, R. ; Bakker, M.G.; Havrilla,
G.; Montoya, V.; Kenik, E.A. ; Shamsuzzoha, M. Microporous Mesoporous
Mater. 2006, 97, 114; (b) Chernysheva, M.V. ; Sapoletova, N.A. ; Eliseev,
A.A. ; Lukashin, A.V. ; Tretyakov, Y.D. ; Goernert, P. Pure and Appl.Chem.
2006, 78, 1749; (c) Zhang, Z.; Dai, S.; Blom, D. A.; Shen, J. Chem.
Mater. 2002, 14, 965; (d) Luo, H.; Wang, D.; He, J.; Lu, Y. J. Phys.
Chem. B 2005, 109, 1919.

\bibitem{6_ Vidal_2009}(a) Vidal, F.; Zheng, Y.; Milano, J.; Demaille,
D.; Schio, P.; Fonda, E.; Vodungbo, B. Appl. Phys. Lett. 2009, 95,
152510 ; (b) Schio, P.; Vidal, F.; Zheng, Y.; Milano, J.; Fonda, E.;
Demaille, D.; Vodungbo, B.; Varalda, J.; de Oliveira, A. J. A. ; Etgens
V. H. Phys. Rev. B 2010, 82, 094436.

\bibitem{7_Li_2006}Li, Y.; Tevaarwerk, E. ; Chang, R.P.H. Chem. Mater.
2006, 18 , 2552.

\bibitem{8_Dumestre_2002}(a) Dumestre, F. ; Chaudret, B. ; Amiens,
C. ; Fromen, M.-C. ; Casanove, M.-J. ; Renaud, P. ; Zurcher, P. Angew.
Chem. Int. Ed. 2002, 41, 4286. (b) Dumestre, F. ; Chaudret, B. ; Amiens,
C. ; Respaud, M. ; Fejes, P. ; Renaud, P. ; Zurcher, P. Angew. Chem.
Int. Ed. 2003, 42, 5213.

\bibitem{9_Soumare_2008}Soumare, Y. ; Piquemal, J.-Y. ; Maurer, T.
; Ott, F. ; Chaboussant, G. ; Falqui, A. ; Viau, G. J. Mater. Chem.
2008, 18, 5696.

\bibitem{10_Xie2012} (a) Y. Xie, J.-M. Zhang, J. Phys. Chem. Sol.
73 (2012) 530-534 ; (b) A. S. Samardak, E. V. Sukovatitsina, A. V.
Ognev, L. A. Chebotkevich, R. Mahmoodi, S. M. Peighambari, M. G. Hosseini,
F. Nasirpouri, J. Phys. : Conf. Ser. 345 (2012) 012011 ; (c) A. Morelos-Gómez,
F. López-Urías, E. Muñoz-Sandoval, C. L. Dennis, R. D. Shull, H. Terrones,
M. Terrones, J. Mater. Chem. 20 (2010) 5906-5914 ; (d) X. Huang, L.
Li, X. Luo, X. Zhu, G. Li, J. Phys. Chem. C 112 (2008) 1468-1472.

\bibitem{11_Pak2011} O. S. Pak, W. Gao, J. Wang, E. Lauga, Soft Matter
7 (2011) 8169-8181.

\bibitem{10_Maurer_2007}Maurer, T.; Ott, F. ; Chaboussant, G. ; Soumare,
Y.; Piquemal, J.-Y. ; Viau, G. Appl. Phys. Lett. 2007, 91, 172501.

\bibitem{11_Soumare}Soumare, Y. ; Garcia, C. ; Maurer, T. ; Chaboussant,
G. ; Ott, F. ; Fiévet, F. ; Piquemal J-Y.; Viau, G. Adv. Funct. Mater.
2009, 19, 1971.

\bibitem{12_Soulantica}Soulantica, K.; Wetz, F.; Maynadié, J.; Falqui,
A.; Tan, R. P.; Blon, T.; Chaudret, B.; Respaud, M. Appl. Phys. Lett.
2009, 95, 152504.

\bibitem{13_Viau}Viau, G.; Garcia, C.; Maurer, T.; Chaboussant, G.;
Ott, F.; Soumare, Y.; Piquemal, J.-Y. Phys. Status Solidi A 2009,
206, 663.

\bibitem{14_Ott}(14) Ott, F.; Maurer, T.; Chaboussant, G.; Soumare,
Y.; Piquemal, J.-Y.; Viau, G. J. Appl. Phys. 2009, 105, 013915.

\bibitem{15_Maurer}Maurer, T.; Zighem, F.; Fang, W.; Ott, F.; Chaboussant,
G.; Soumare, Y.; Ait Atmane, K.; Piquemal, J.-Y.; Viau, G. J. Appl.
Phys. 2011, 110, 123924.

\bibitem{16_Maurer} Maurer, T.; Zighem, F.; Ott, F.; Chaboussant,
G.; André, G.; Soumare, Y.; Piquemal, J.-Y.; Viau, G.; Gatel, C. Phys.
Rev. B 2009, 80, 064427.

\bibitem{17_Karim}Karim, S; Toimil-Molares, M.E.; Balogh, A.G.; Ensinger,
W.; Cornelius, T.W.; Khan, E.U.; Neumann, R. Nanotechnology 2006,
17, 5954

\bibitem{18_Nisoli} Nisoli, C.; Abraham, D.; Lookman, T.; Saxena,
A. Phys. Rev. Lett. 2009, 102, 245504.

\bibitem{19_Ciuculescu}Ciuculescu, D.; Dumestre, F.; Comesaña-Hermo,
M.; Chaudret, B.; Spasova, M.; Farle, M.; Amiens, C., Chem. Mater.
2009, 21, 3987.

\bibitem{22_Ono }Ono, F. J. Phys. Soc. Jap. 1981, 50, 2564.

\bibitem{23_Barnier1961} Y. Barnier, R. Pauthenet, G. Rimet, C. R.
Hebd. Acad. Sci. 253 (1961) 400.

\bibitem{24_Handley} R. C. O'Handley, Modern Magnetic Materials,
Principles and Applications, Wiley, New York, 2000, p. 319.\end{thebibliography}
\end{document}